\documentclass[
nopreprint,%choose either nopreprint or preprint
showpacs,preprintnumbers
]{jasatex}

\usepackage{graphicx}% Include figure files
\usepackage{dcolumn}% Align table columns on decimal point
\usepackage{amsmath,amsfonts}% popular packages from the American Mathematical Society
\usepackage{bm}% Bold Math package
\newcommand{\myint}[0]{\int \limits}
\newcommand{\intinfty}[0]{\int \limits_{-\infty}^{\infty}}

\newcommand{\Si}[0]{\mbox{Si}}
\newcommand{\Ci}[0]{\mbox{Ci}}

\begin{document}

\preprint{}

\title[Bending Wavelet]{\center Bending Wavelet for Flexural Impulse Response}

\author{Richard B\"ussow}
 \altaffiliation{Institute of Fluid Mechanics and Engineering Acoustics, Berlin University of Technology}%Lines break automatically or can be forced with \\
\affiliation{Einsteinufer 25, 10587 Berlin}%Line breaks may be forced with \\ here, too

\author{Richard B\"ussow}
 \homepage{http://www.tu-berlin.de/fb6/ita}

\date{\today}% It is always \today, today,

\begin{abstract}
The work addresses the definition of a wavelet  that is adapted to analyse  a
flexural impulse response. The wavelet gives the  opportunity to directly
analyse the dispersion characteristics of a pulse. The aim is to localize a  
source or to measure material parameters. An overview of the mathematical 
properties of the wavelet is presented. An algorithm to extract the dispersion 
characteristics with the use of genetic algorithms is outlined. The application 
of the wavelet is shown in an example and experiment. 
\end{abstract}



\maketitle
\section{Introduction}
 The Morlet wavelet transform is a popular method for
 time-frequency analysis. Its application for  acoustic signals can be found in several publications. These 
 publications deal for example with the analysis of dispersive waves
 \cite{onsay,kishimoto},  source or damage localization
 \cite{yamada,gaulident,rucka,junsheng,railroad,messina},  investigation of
 system parameters \cite{hayashi,ta} or active control \cite{berry98wavelet}. 
 A comparison of  the short time Fourier transform and the Morlet wavelet
 transform is done by Kim
 et.al.  \cite{kimkim}.  It is found that the continuous wavelet transform CWT
 of acoustic signals is a promising method to obtain the time - frequency
 energy  distribution of a signal.\\
These applications are based on the evaluation of the frequency
dependent  arrival time of a pulse in dispersive media. The underlying concept
of  this method will be briefly explained for a one-dimensional structure (e.g. a beam). \\
A fundamental difference between most waveforms in structures and fluids is the 
dispersion. A pulse propagating in a structure with the frequency dependent
group  velocity $c_g$ changes its shape. Due to this dispersion the pulse is not
 recognizable with correlation techniques that can be useful for locating
 airborne  sound sources \cite{micarray}. \\
The wavelet transform is very useful to extract exactly the arrival time $t_a$ 
of an pulse in a dispersive media
\begin{equation}
t_a = x / c_g.
\end{equation}
The continuous wavelet transform $W_\psi^y$ of a function $y$ is 
\begin{equation}
W_\psi^y(a,b) = \frac{1}{\sqrt{c_\psi |a|}} \intinfty y(t) \: \psi\left(\frac{t - b}{a} \right) \: dt.
\label{cwtdef}
\end{equation}
The analogue to the Fourier transforms spectrogram is the scalogram defined  as
$|W_\psi^y (a,b)|^2$.
It can be shown that for a fixed scaling parameter $a$ the arrival  time $t_a$
is the point in time where the maximum of the scalogram occurs 
\begin{equation}
max\left( |W_\psi^y (a,b)|^2\right) = |W_\psi^y (a,b=t_a)|^2.
\end{equation}
To locate a source one needs 
\begin{enumerate}
\item the point in time the pulse occurred, the group velocity and a sensor, or 
\item two sensors, or
\item one sensor measuring two distinguishable wave types \cite{jiao}.
\end{enumerate}
If the position of the source is known, it is possible to extract material
parameters  \cite{hayashi,ta}.\\
To improve this method, dispersion based transforms have been proposed
\cite{yoonkim,liu},  which is based on a method called  Chirplet transform
\cite{mannVI91,mannsp}. These transforms improved the analysis. Nevertheless the
bending wavelet that is presented is a new approach. \\
Here a different wavelet suitable for bending waves which can be modelled with the Euler beam
 theory is presented.\\
  The underlying concept is not to measure
the arrival time but to extract directly the dispersion of the pulse. The
dispersion  of the pulse is dependent on the distance between source and
receiver and the material properties.  If it is possible to extract exactly the
spreading of the pulse one has  directly the distance or the material
properties, depending on which is known. To define a wavelet that extracts the 
dispersion characteristics it is necessary to know the impulse response
function  in the time domain. For plates which can be modelled with the Euler beam
 theory this function is derived first by Boussinesq\cite{boussinesq} and can be
 found in textbooks\cite{heckl}. For beams it is treated in
a companion publication\cite{bendimpulse} since only the Green's functions for 
a initial deflection and velocity are found in the literature\cite{nowacki,meirovitch}.\\
The velocity $v(r,t)$ resulting from the bending wave propagation on a
infinite plate  of a  force impulse $F_a(t) = F_0 \: \delta(t)$ at $r=0$ is
 \begin{equation}
v(r,t) = \frac{\hat{F}_0}{4 \pi t \sqrt{ B' m''}} \sin \left( \frac{r^2}{4 \zeta t} \right),
\label{schnelleplatte}
\end{equation}
where $r$ is the distance from the source, $\zeta = \sqrt{B'/m''}$, bending
stiffness $B'=E h^3/(12(1 - \nu^2))$, $E$ the elastic or Young's modulus, $h$ the plate
thickness, $\nu$ the Poisson's ratio and $m''$ the mass per unit area. \\
The bending wave velocity $v(x,t)$ on a infinite beam, resulting of a force impulse $F_a(t) = F_0 \: \delta(t)$ at $x=0$ is given by
\begin{equation}
v(x,t) = \frac{F_0 H(t)}{ m'}  \sqrt{ \frac{2 }{\pi \zeta t}} \cos \left( 
\frac{x^2}{4 \zeta t} \right) \: .
\label{vdirac}
\end{equation}
therein $x$ is the distance from the source, $\zeta=\sqrt{B/m'}$, where $B$ is
the bending stiffness of the beam and $m'$ mass per unit length.\\
The term 
\begin{equation}
d_i=\frac{x^2}{4 \zeta}.
\label{di}
\end{equation}
is the factor that controls this spreading and is called dispersion factor.
Whereas the dispersion factor is a time value the nondimensional term  
\begin{equation}
Di=\frac{x^2 f_{max}}{4 \zeta},
\label{Di}
\end{equation}
is called dispersion number. The applicability of the following method depends on
the dispersion number. Higher dispersion numbers result in a longer useful time
period and this case is better to analyse. An exact quantification
is given in the following, equation (\ref{nminmax}). A high dispersion number is
the reason for choosing a thin plate and a slender beam. \\
In the following a new adapted wavelet will be derived to extract the dispersion
 factor from the measured pulse. Usually a wavelet is designed to localize a
 certain  frequency. In contrary the proposed wavelet localizes a frequency
 range  that is distributed over the wavelet length just like equation
 (\ref{schnelleplatte}) or (\ref{vdirac}). Such a choice follows the paradigm of
 signal processing, that "the analysing function should look like the signal".\\
One may interpret the continuous Wavelet transform as a cross-correlation of 
$y$ and $\psi$. Hence, the idea is to find the function
which is highly 
correlated with the impulse response. The difference is the role of the scaling 
parameter $a$. It is vital to produce the presented results to use the scaling 
parameter as it is defined in equation (\ref{cwtdef}). \\
The dispersion factor is determined by the scaling factor with the highest value of
the scalogram. In principle this can be done with a fine grid of $(a,b)$
values. A  more efficient way is to use an optimisation scheme. Gradient based 
optimisation is not reliable in finding a global optimum. A second problem is
the  localisation of several overlapping pulses. A well known method that is able 
to fulfill these requirements are genetic algorithms.  
\section{Bending wavelet}
Several different definitions based on the Morlet wavelet and the Chirplet 
transform\cite{mannVI91,mannsp} have been investigated. For brevity an extensive
 discussion about the different efforts is omitted. The details of the
 mathematical  background of the wavelet transform can be found in the
 literature  \cite{Mallat98wavelet, wavelets}. \\
The section begins with the definition of a wavelet with compact support and
zero-mean.  It follows a comment on the amplitude and frequency distribution
and ends  with possible optional definitions.
\subsection{Definition}
\begin{figure}[ht]
\includegraphics[width=0.4 \textwidth]{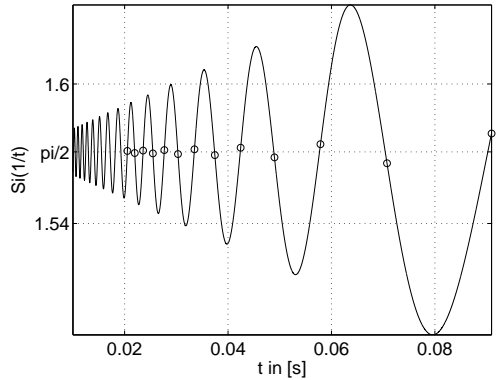}
\caption{$\Si(1/t)$ and circles at $1/t = (2n - 1) \pi/2$}
\label{picsi}
\end{figure}

The mother wavelet  
\begin{equation}
\psi_s(t) = \left\{ \begin{array}{ll} \frac{\sin(1/t)}{t} & \mbox{ for } t_{min} < t < t_{max} \\ 0 & \mbox{ otherwise} \end{array}\right. ,
\label{biegewav}
\end{equation}
is called bending wavelet.\\
A wavelet $\psi$ must fulfill the admissibility condition 
\begin{equation}
0 < c_{\psi} = 2 \pi \intinfty \frac{|\hat{\psi}(\omega)|^2}{|\omega|} \: d\omega < \infty,
\end{equation}
where $\hat{\psi}(\omega)$ is the Fourier transform of the wavelet. The
proposed wavelet (\ref{biegewav}) has  a compact support $(t_{min},t_{max})$, which means that
the  admissibility condition is fulfilled if 
\begin{equation}
\myint_{t_{min}}^{t_{max}} \psi (t) \: dt = 0
\label{mittelwertfrei} 
\end{equation}
holds. To fulfill the admissibility condition $t_{min}$ and
$t_{max}$ are defined so, that equation (\ref{mittelwertfrei}) holds.  With the integral - sine function $\Si(x) = \int_0^x \sin(t)/t \: dt$ one finds that
\begin{equation}
\myint_{t_{min}}^{t_{max}} \frac{\sin(1/t)}{t} \: dt =  \Si \left( \frac{1}{t_{max}} \right) - \Si \left( \frac{1}{t_{min}} \right).
\end{equation}
Since $\lim_{t \to 0} \Si(1/t) = \pi/2$ and that the $\Si$-function for $t<2/\pi$ oscillates around $\pi/2$, one is able to chose $t_{min}$ and $t_{max}$ so, that 
\begin{equation}
\Si(1/t_{min}) = \Si(1/t_{max})= \pi/2.
\label{sibedingung}
\end{equation}
This is a very easy option to define a wavelet and it will be used later to
define  similar wavelets. Equation (\ref{sibedingung}) can only be solved
numerically  so a good approximation should be used that leads to a simple  
expression for $c_{\psi}$. The support of the wavelet is defined by
\begin{equation}
\frac{1}{t_{min/max}} = \frac{(2n_{max/min} - 1) \pi}{2}.
\label{tminmax} 
\end{equation}
The function $\Si(1/t)$ and proposed possible values of $t_{min}$ or $t_{max}$
are  plotted in figure \ref{picsi}. In the worst case for $t_{max} = 2/\pi$ and 
$t_{min} \rightarrow \infty$ the difference in equation (\ref{mittelwertfrei})
is  around $0.2$, but for higher values of $n_{min}$ the magnitude is in the 
order of other inaccuracies, so that it should be negligible. Like the Morlet 
wavelet, which fulfills the admissibility condition in a asymptotic sense the
bending wavelet fulfills the admissibility for $\lim t \rightarrow 0$. %
The value of the constant is calculated $c_{\psi}$ with the norm in the Lebesque space $L^2$ of square integrable functions 
\begin{equation}
||\psi(t)||_2 = \left( \intinfty \psi(t)^2 \: dt \right)^{1/2}.
\label{psil2}
\end{equation}
The integral in equation (\ref{psil2}) is 
\begin{equation}
\begin{array}{l}
\myint_{t_{min}}^{t_{max}} \sin(1/t)^2/t^2 \: dt = \\
\frac{1}{4} \left( \frac{2}{t_{min}} - \frac{2}{t_{max}} - \sin \left( \frac{2}{t_{min}} \right) + \sin \left( \frac{2}{t_{max}} \right) \right).
\end{array}
\end{equation}
With the proposed choice of $t_{min}$ and $t_{max}$, the sine vanishes and a
normalised $||\psi(t)||_2 = 1$ wavelet is obtained if $c_\psi$ is chosen to be
\begin{equation}
c_\psi = \frac{1}{2} \left( \frac{1}{t_{min}} - \frac{1}{t_{max}} \right) = \frac{\pi}{2}(n_{max} - n_{min}).
\end{equation}

\subsection{Displacement-invariant definition}
Wavelets that are defined by real functions have the property that the
scalogram  depends on the phase of the analysed function. Wavelets that are
complex  functions like e.g. the Morlet wavelet are called
displacement-invariant.  A wavelet $\psi = \psi_c + i \psi_s$ that consists of a
 sine, $\psi_s$ equation (\ref{biegewav}), and a cosine wavelet which is 
\begin{equation}
\psi_c(t) = \left\{ \begin{array}{ll} \frac{\cos(1/t)}{t} & \mbox{ for } t_{min} < t < t_{max} \\ 0 & \mbox{ otherwise} \end{array}\right..
\label{biegecoswav}
\end{equation}
can be beneficial. With the integral - cosine function $\Ci(x) = \int_0^x \cos(t)/t \: dt$ one finds that
\begin{equation}
\myint_{t_{min}}^{t_{max}} \frac{\cos(1/t)}{t} \: dt =  \Ci \left( \frac{1}{t_{max}} \right) - \Ci \left( \frac{1}{t_{min}} \right).
\end{equation}
The analogous definition of the value $\pi/2$ for the $\Ci$-function is  
\begin{equation}
\Ci(1/t_{min}) = \Ci(1/t_{max})= 0.
\label{cibedingung}
\end{equation}
The approximation is given by
\begin{equation}
\frac{1}{t_{min/max}} = n_{max/min} \pi.
\label{citminmax} 
\end{equation}
The effect is that the real- and the imaginary part of the resulting wavelet do not share the same support. This is an awkward definition of a mother wavelet but the difference between the two supports is rather small if the same value for $n_{max/min}$ is used. To keep things simple only the real valued sine wavelet is used in following.

\subsection{Orthogonality of the bending wavelet}
\label{secortho}
\begin{figure}[ht]
\includegraphics[width=0.4 \textwidth]{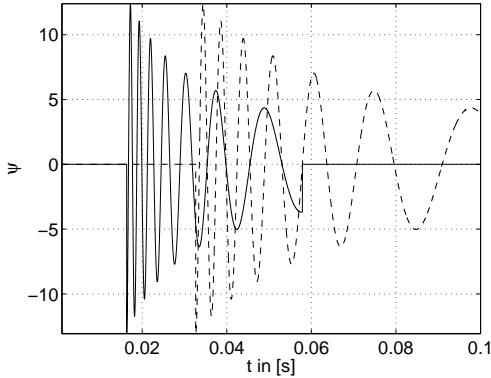}
\caption{Bending wavelet (\ref{biegewav}) for $a_1=1$ (solid) and $a_2 = 2$ (dashed), $n_{min} = 4$ and $n_{max} = 12$}
\label{zweibwavs}
\end{figure}

The trigonometric functions that are used for the Fourier transform establish
an orthogonal base.  Hence, the Fourier transform has the convenient
characteristic that  only one value represents one frequency in the analysed
signal.  Every deviation of this is due to the windowing function that is
analysed  with the signal. Already the short time Fourier transform is not
orthogonal,  if the different windows overlap each other. Because of this
overlap the  continuous wavelet transform can not be orthogonal. The proposed
wavelet  should still be investigated since it is instructive for the
interpretation  of the results. \\ 
The condition for an orthogonal basis in Lebesgue space $L^2$ of square integrable functions is
\begin{equation}
(\psi_j,\psi_k) = \intinfty \psi_j \psi_k = \delta_{jk}. 
\label{orthonormal}
\end{equation}
Two different wavelets $\psi_j$ and $\psi_k$ can be obtained by using different
scaling parameters $a$ and/or different displacement parameters $b$. Here the effect of two scaling parameters is investigated, so the following integral is to be solved
\begin{equation}
\intinfty a_j a_k \sin(a_j/t) \sin(a_k/t) / t^2 \: dt . 
\end{equation}
To illustrate the integral, two different versions of the wavelet are plotted
in figure \ref{zweibwavs}. 

One finds that
\begin{equation}
\begin{array}{l}
\intinfty a_j a_k \sin(a_j/t) \sin(a_k/t) / t^2 \: dt =  \\
\left. \frac{a_j a_k }{a_k^2 - a_j^2} \left( a_j \cos \left( \frac{a_j}{t} \right) \sin \left( \frac{a_k}{t} \right) - a_k \cos \left( \frac{a_k}{t} \right) \sin \left( \frac{a_j}{t} \right) \, \right) \right|_{t_{min}}^{t_{max}}.
\end{array}
\label{l2integral}
\end{equation}
The sine term in equation (\ref{l2integral}) vanishes since $t_{min/max}$, also scale with $a$, but actually there are two different values of $a$ and so not all four sine terms vanish. With this result one expects a rather broad area in $(a,b)$ with high values of the scalogram.

\subsection{Time amplitude/frequency distribution}
The time frequency distribution of the wavelet for a certain scaling factor $a$ is determined by the argument of the sine function. The instaneous frequency $\omega(t)$ can theoretical be obtained with a relationship for almost periodic functions, see Bochner \cite{bochnerfp}. The actual frequency of the wavelet is 
given by
\begin{equation}
\sin \varphi(t) \rightarrow \omega(t) = \varphi'(t) = \frac{a}{t^2}.
\label{omegasin}
\end{equation}
The $1/t$ leading term affects the amplitude of the wavelet. Usually it is desired that the whole signal contributes linearly to the transform. To achieve this it is useful to have an amplitude distribution over time of the wavelet that is reciprocal to the amplitude distribution of the analysed signal. Since the bending wavelet has the same amplitude distribution as equation (\ref{schnelleplatte}) this may lead to stronger weighting of the early high frequency components of the impulse response. \\
A force impulse that compensates the amplitude distribution of the impulse response follows a $\sim 1/\omega^2$ dependence.

\section{Continuous wavelet transform with the bending wavelet}
The application in the given context is to extract precisely the scaling factor
with the highest  value. How this is achieved will be discussed in the next
section.  Here the realisation of a transform with a set of scaling factors is
presented since  it is illustrative. \\
The algorithm implementing the continuous wavelet transform with the bending
wavelet  can not be the same as the algorithm implementing a transform with any 
continuous wavelet, like the Morlet wavelet. The bending wavelet has a compact
support,  which must be defined prior to the transform. This can be done with a 
estimation of the frequency range and the dispersion number. With the equations 
(\ref{tminmax}) and (\ref{omegasin}) it holds that 
\begin{eqnarray}
n_{max} &=& \mbox{floor} \left( \sqrt{\frac{d_i f_{max}}{ 2 \pi}} + \frac{1}{2} \right) \nonumber \\
n_{min} &=& \mbox{ceil} \left( \sqrt{\frac{d_i f_{min}}{2 \pi}} + \frac{1}{2} \right),
\label{nminmax}
\end{eqnarray}
where floor$(\cdot)$ rounds down towards the nearest integer and ceil$(\cdot)$
rounds up.  The knowledge of a useful frequency range should not provide any
problems. But to have  to know beforehand which dispersion number will dominate
the  result is rather unsatisfactory. A more practical solution is to calculate
the  corresponding $n$-value within the algorithm, which is an easy task since
$a=d_i$.  The problem with this possibility is that the support of the wavelet 
changes within the transform. Since the support is part of the wavelet this
means  that strictly one compares the results of two different wavelets. Since
the  wavelet is normalised the effect is rather small, but nevertheless it
should  be interpreted with care.

\subsection{Example}
\begin{figure}[ht]
\includegraphics[width=0.4 \textwidth]{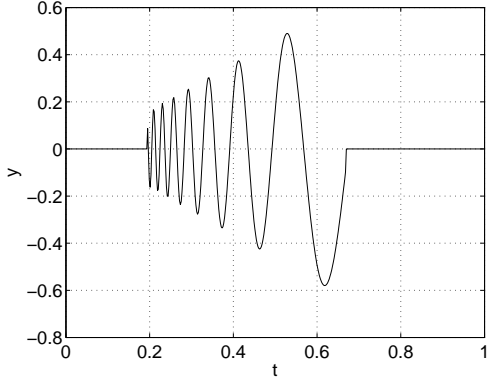}
\caption{Analysed example function (\ref{excwtbend})}
\label{bwavtestfunc}
\end{figure}

\begin{figure}[ht]
\includegraphics[width=0.4 \textwidth]{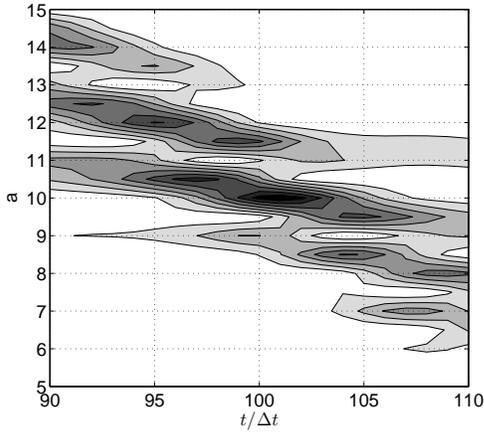}
\caption{Contour plot of the scalogram build with the bending wavelet transform of equation (\ref{excwtbend}) with $f_{max}=f_s/8$ and $f_{min}=4$}
\label{contbwav}
\end{figure}

\begin{figure}[ht]
\includegraphics[width=0.4 \textwidth]{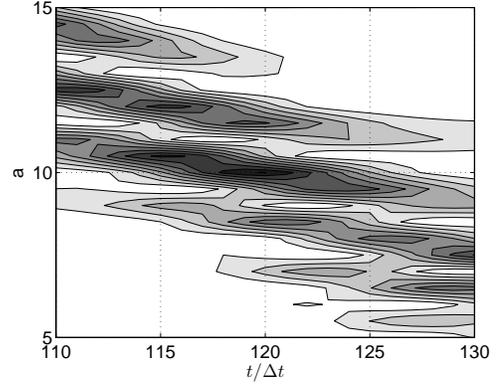}
\caption{Contour plot of the scalogram build with the bending wavelet transform of equation (\ref{excwtbend}) with $f_{max}=f_s/12$ and $f_{min}=4$}
\label{contbwavfs12}
\end{figure}

To illustrate the use of the proposed wavelet the following function is transformed 
\begin{equation}
y(t) = \left\{
\begin{array}{ll}
 t \sin(a/t), & \mbox{ for } t_{min} < t < t_{max} \\
 0 & \mbox{otherwise,}
\end{array} \right.
\label{excwtbend}
\end{equation}
with $a=10$. The sampling frequency is $f_s = 2^{9}$, $t_{min}$ and $t_{max}$
are  defined by the corresponding values of $f_{max} = f_s / 8$ and $f_{min} =
4$,  for convenience the point $t=t_{min}$ is shifted to $10 \Delta t$. The
example function in equation (\ref{excwtbend}) is plotted in figure \ref{bwavtestfunc}. \\
The example function is transformed with the algorithm that calculates 
$n_{min}$ and $n_{max}$  with the corresponding value of $f_{max}$, $f_{min}$
and $a$. The choice of the frequency range is critical. If it is too small,
information will be  lost and if it is too big the parts that overlap the pulse
may distort  the result. Here the same frequency range as the analysed function
is used.\\
The resulting scalogram is not plotted directly against the factor $b$, but
shifted with  the value of $t_{min}$. This means that the maximum value is at $10
\Delta t$ 
(figure \ref{contbwav}), which is the value of $t$ where $f_{max}$ is located. 
The maximum value is shifted when $f_{max} = f_s/12$ is used, as can be seen in 
figure \ref{contbwavfs12}. \\

One may recognise that there are very high values if the wavelet is shifted and
scaled along  the curve $a/t$. This is expected theoretically, as discussed in section
\ref{secortho} and can be interpreted descriptive since the wavelet does not
localize one  frequency, but has a wide frequency range that spreads over time.
It can be  quantified with equation (\ref{orthonormal}). Evaluating this
integral numerically for the values of $a_1=10$, $a_2=11.75$ and $b_2 = 7
\Delta t$ results  in value of $0.68$, which means that the peak at $a_2=11.75$
has $68\%$  of the peak at $a_1=10$. \\
 This problem of non-orthogonality is addressed by the following algorithm. The
 pulse  is extracted from the signal by first locating the position $t_{start}$
 in the  signal, where $f_{max}$ has its maximum. This is done by a Morlet
 wavelet  transform with which one may find the value of $t_{start}$ that has
 the highest  value of $f_{max}$. Now the transformation with the bending
 wavelet is only  done in the vicinity of $t_{start}$. Technically the
 displacement  parameters $b$ are defined with $t_{start}$.  

\section{Localization with a Genetic Algorithm} 
Genetic algorithms (GA) form a particular class of evolutionary algorithms that
use  techniques inspired by evolutionary biology such as inheritance, mutation, 
selection, and crossover. Genetic algorithms are categorized as global search  
heuristics. Details of the method can be found in the extensive literature, 
e.g.\cite{pohlheim}. \\
The genetic algorithm is chosen, since it is usually very reliable in finding a 
global optimum and its ability to find Pareto optimums to locate several
pulses.  However the drawback is the slow convergence, that can be improved with
 a local search method. Recent publications on this topic are \cite{gageo,gacarin}.
 \\
The implementation is done with functions provided by the open source  Matlab
toolbox \cite{gatoolbox}, if not stated otherwise. Principally possible but not 
used in the example is the localization of two pulses that are overlapping. For 
the sake of brevity a discussion on how this can be achieved will be omitted.   
The algorithm works with two variables, the displacement parameter and the 
scaling factor  of the bending wavelet. \\
The displacement parameter is defined by $t_{start} \pm \lambda/2$ or smaller  
values. This depends on the size of the Morlet wavelet. For discrete functions
it  is an integer value but nevertheless implemented as a floating point number,
 because of lacking support for such a combination in the toolbox. This fact is 
 taken into account when calculating the wavelet transform. 
A pseudo-code that describes the genetic algorithm can be found in the appendix \ref{gacode}.\\
In the end only the fittest individual is extracted. The algorithm is usually
quite reliable.  Since it is a stochastic method, it can be beneficial to
restart  the whole process or to work with several sub-populations.
\subsection{Example}
\begin{figure}[ht]
\includegraphics[width=0.4 \textwidth]{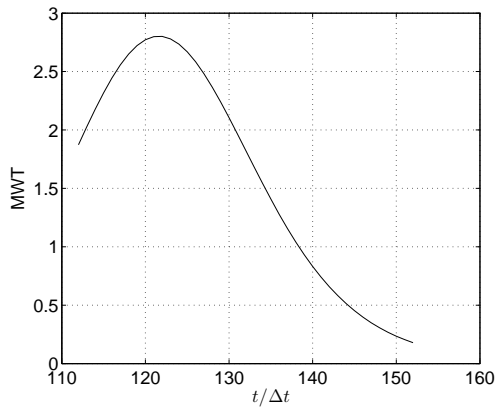}
\caption{Morlet wavelet transform of equation (\ref{excwtbend}) with $f=f_s/12$}
\label{tstart}
\end{figure}

\begin{figure}[ht]
\includegraphics[width=0.4 \textwidth]{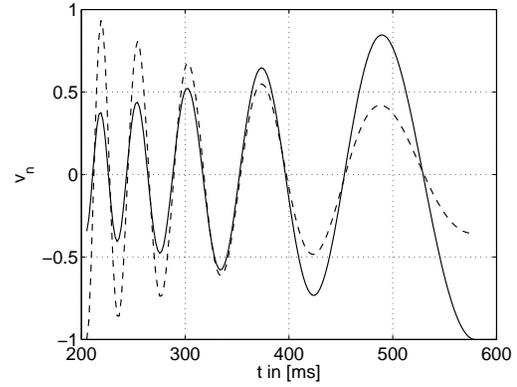}
\caption{Bending wavelet at $a=9.99$ and the example function}
\label{funcga}
\end{figure}

As an example, the already transformed equation (\ref{excwtbend}) is
investigated.  The frequency range is the same as the example plotted in figure
\ref{contbwavfs12}.  As a first step the value $t_{start}$ is calculated with  a
Morlet wavelet transform at $f_s/12$. The result is plotted in figure
\ref{tstart}, where the maximum is at $t_i =122$. From figure \ref{contbwavfs12} one 
may conclude that the correct value is $120$ this slight deviation is due to the
 fact that the Morlet wavelet has a rather broad frequency resolution and the  
 amplitude of the signal is increasing with time. If the signal $\sin(a/t)/t$  
 is used, the maximum is located at $118$. The number of individuals is chosen  
 to $100$ and the number of generations to $400$.\\
The initial chromosome has a time index range of $(106,138)$ which is $3/4
\lambda$  and the range of the scaling factor is $(5,15)$. The obtained scaling 
factor is  $a=9.99$. The frequency range of the bending wavelet is shorter than 
the values $f_{wmax}=37.7<42.7=f_s/12$ and $f_{wmin}=4.756<4<f_{min}$, this due 
to equation \ref{nminmax}. The wavelet with the best scaling factor and the  
example function are plotted together in figure \ref{funcga}. One may recognize 
that the frequency-time distribution of both function match.

\subsection{Experimental results}
Experiments on a beam and a plate are discussed in detail in combination with
the impulse  response functions\cite{bendimpulse}. The method was used to
extract the  dispersion factor $d_i$. The theoretical functions
(\ref{schnelleplatte})  and (\ref{vdirac}) with the obtained dispersion factor
$d_i$  are compared with the measured curves. A good agreement of theory and  
experiment shows the applicability of the presented transform.

\subsection{Comparision with Morlet Wavelet}
The drawback of the proposed method is that it is only applicable to bending
waves that can be modelled with the simple Euler beam theory or any wave that 
has a dispersion relation that follows a $\sim 1/\sqrt{\omega}$  dependence.
Another precondition is the rather high dispersion number. 
Nevertheless bending waves are dominant in structure
borne sound problems and for thin structures the simplifications of the Euler
bending theory are usually valid. \\ 
The method is accurate, fast and easy to implement. In the
experiments a  source could be localized with a deviation lower than $10\%$. One
can build real time applications for source detection. The method has
also principle advantages, it is possible to
\begin{enumerate}
\item obtain the distance of a impulse or the material properties with only one measurement, 
\item analyse two overlapping pulse, which is not possible with the maximum of
the Morlet  wavelet transform where for each frequency one maximum value is extracted.
\end{enumerate}

\section{Concluding Remarks}
A definition of a new adapted wavelet, the bending wavelet, is given. The
mathematical properties of the bending wavelet are discussed. It is shown in
examples that the transform is useful to analyse a flexural impulse response. \\
Besides source localisation a possible application is the measurement of material
properties in a built-in situation of finite structures, since the method does not depend
on the boundary conditions. \\
The choice of a useful frequency range can be problematic. It may be useful to
first  analyse the signal with a Morlet wavelet transform to find a useful range.  
\begin{appendix}
\section{Appendix}
\label{gacode}
\begin{verbatim}
1 Linear distributed 
  initial chromosome of t and a.
2 Bending wavelet transform with 
  the initial chromosome.
3 Assignment to the 
  current population. 
Repeat
    4 Self-written rank based 
      fitness assignment (current pop.).
    5 Selection with 
      stochastic universal sampling.
    6 Recombination with the 
      extended intermediate function.
    7 Real-value mutation with 
      breeder genetic algorithm. 
    8 Bending wavelet transform with the 
      selected chromosome.
    9 Self-written fitness based insertion
      with 70% new individuals.
Until max number of generations 
\end{verbatim}
\end{appendix}
\bibliographystyle{elsart-num}
\bibliography{lite}

\end{document}